\documentclass[%
 reprint,doublecolumn, notitlepage,
 amsmath,amssymb,
 aps,
 superscriptaddress,
]{revtex4-2}

\makeatletter
\DeclareRobustCommand{\change}{%
  \@bsphack
  \leavevmode
  \color{red}%
  \@esphack
}
\usepackage{hyperref}
\hypersetup{
    colorlinks=true, 
    linkcolor=cyan,
    citecolor=magenta, 
    filecolor=magenta, 
    urlcolor=cyan,
    runcolor=cyan
}
\usepackage[capitalise]{cleveref}

\usepackage{newtxtext}
\usepackage{newtxmath}

\usepackage{braket}%Dirac Notation in QM
\usepackage{graphicx}% Include figure files
\usepackage{dcolumn}% Align table columns on decimal point
\usepackage{bm}% bold math
\usepackage{color}
\usepackage{tabularx}
\usepackage{bbm}
\newcommand{\bes} {\begin{subequations}}
\newcommand{\ees} {\end{subequations}}
\newcommand{\ketb}[2]{|{#1}\>\!\<#2|}
\newcommand{\mc}{\mathcal}
\def\>{\rangle}
\def\<{\langle}
\def\Tr{\mathrm{Tr}}

\begin{document}

%\preprint{}

\title{Error budget of parametric resonance entangling gate with a tunable coupler}

\author{Eyob A. Sete}
\thanks{Corresponding author, eyob@rigetti.com}
\affiliation{%
Rigetti Computing, 775 Heinz Avenue, Berkeley, California 94710, USA
}%
\author{Vinay Tripathi}
\affiliation{%
Department of Physics \& Astronomy, and Center for Quantum Information Science \& Technology, University of Southern California, Los Angeles, California 90089, USA
}%
\author{Joseph A. Valery}
\affiliation{%
Rigetti Computing, 775 Heinz Avenue, Berkeley, California 94710, USA
}

\author{Daniel Lidar}
\affiliation{%
Departments of Electrical \& Computer Engineering, Chemistry, Physics \& Astronomy, and Center for Quantum Information Science \& Technology, University of Southern California, Los Angeles, California 90089, USA
}%

\author{Josh Y. Mutus}
\affiliation{%
Rigetti Computing, 775 Heinz Avenue, Berkeley, California 94710, USA
}%

\date{\today}% It is always \today, today,

\begin{abstract}
We analyze the experimental error budget of parametric resonance gates in a tunable coupler architecture. We identify and characterize various sources of errors, including incoherent, leakage, amplitude, and phase errors. By varying the two-qubit gate time,  we explore the dynamics of these errors and their impact on the gate fidelity. To accurately capture the impact of incoherent errors on gate fidelity, we measure the coherence times of qubits under gate operating conditions. Our findings reveal that the incoherent errors, mainly arising from qubit relaxation and dephasing due to white noise, limit the fidelity of the two-qubit gates. Moreover, we demonstrate that leakage to noncomputational states is the second largest contributor to the two-qubit gates infidelity, as characterized using leakage-randomized benchmarking. The error budgeting methodology we developed here can be effectively applied to other types of gate implementations.
\end{abstract}

\pacs{Valid PACS appear here}% PACS, the Physics and Astronomy
\maketitle
\section{Introduction}

Superconducting qubits have a pivotal role in the advancement of quantum computing, as evidenced by recent developments in the field~\cite{Kandala:2017aa, Minev2019, Arute:2019aa, Arute2020, wu2021strong,Krinner2022, Acharya2023QEC, Sivak2023,pokharel2022demonstration}. A fundamental necessity for quantum computers is the implementation of high-fidelity entangling gates. For superconducting qubits, these typically entail a two-qubit gate reported in various implementations of these gates~\cite{Foxen2020,Mitchell2021,Kandala2021}, achieving fidelities that approach the threshold needed for fault tolerance ~\cite{Hong2020,Foxen2020,sung2020,gold2021,Irfan2022,Ding2023}. These implementations utilize pairs of coupled qubits, where the coupling between the two qubits can be tuned either by modifying the qubits' frequencies~\cite{Barends2014} or coupling strength between elements~\cite{Yan2018, Sung2021}. In setups where this coupling is fixed, microwave pulses can be used to initiate entanglement~\cite{Groot2010, Rigetti2010, Chow2012, Krinner2020}.

Another approach to realize entangling operations involves parametric modulation~\cite{Yurke1989, Tian_2008}, which relies on the modulation of system parameters such as qubit frequencies or coupling strengths to selectively control the interaction between qubits. Typically parametric modulation generates a sideband that can be used to control the interaction between two qubits~\cite{Didier_2017, Caldwell2018, Reagor2018, Abrams2020, Hong2020, Didier19, Valery2022}. A recent advancement utilizes the central band of the modulated qubit to operate the two-qubit gates, known as the \textit{parametric-resonance} gate~\cite{Sete_para_2021}. This approach marks a significant improvement over conventional methods, as it avoids the effective reduction in coupling rates caused by renormalization in sideband interactions. Moreover, parametric resonance gates are insensitive to phase effects during the rise and fall time of the flux pulse~\cite{Abrams2020} because the entangling phase is zeroed out for the central band.
 
Two-qubit gate implementations have increasingly relied on tunable couplers~\cite{Yan2018,Sung2021,Mundada2019,Sete_para_2021, Field2023} to enhance control over qubit interactions. This approach provides the dual advantage of suppressing always-on coupling when the gate is inactive and enabling faster gate operations. Ref.~\cite{Sete_floating_2021} introduced floating tunable couplers, which not only support the conventional configuration with the coupler's frequency higher than the qubits' frequencies but also a novel regime where the coupler's frequency can be lower than that of both qubits. These two operational modes correspond to the \emph{asymmetric} and \emph{symmetric} configurations, respectively. The symmetric coupler regime, in particular, holds substantial potential for scaling up quantum computing systems, as it helps avoid frequency crowding, a common issue when both the coupler and readout resonator frequencies are greater than those of the qubits. 

In this work, we experimentally explore the error dynamics of the parametric resonance interaction and provide a complete error budget to assess the fidelity and performance of two-qubit gates.
We specifically focus on controlled-Z (CZ) gates realized between flux-tunable qubits coupled via a floating tunable coupler in a symmetric configuration, where the coupler frequency is set below that of the qubits. Our study investigates various error channels, dissecting both incoherent and coherent sources. In the incoherent domain, we consider factors including qubit relaxation, Markovian noise-induced dephasing, and dephasing due to non-Markovian $1/\mathrm{f}$ flux noise. On the coherent front, we characterize amplitude errors, deviations in conditional phase rotations (either over or under rotation), and leakage errors. This analysis enables a holistic understanding of the error budget of the parametric resonance gate. 

To analyze the experimental dependence on gate time, we conducted a comprehensive analysis of the error budget of the CZ gate, varying the gate duration to probe how different errors manifest. Our analysis indicates that incoherent errors, notably Markovian noise-induced dephasing, and qubit relaxation, are the primary error sources that limit gate fidelities in our devices, accounting for approximately $83\%$ of the total error. Leakage to noncomputational states emerges as the second most significant factor. Other contributors, including $1/\mathrm{f}$ flux noise-induced dephasing, amplitude variations, and phase errors, collectively constitute less than $2\%$ of the total. The error rates measured in our experiments align with the estimated total errors from incoherent and coherent sources. We derive practical analytical expressions for incoherent errors to obtain these estimates, particularly useful in scenarios where the gate operation time is significantly shorter than the qubits' coherence times. Our formalism applies to both parametric resonance gates and other entangling gate implementations \cite{collodo2020a,xu2020}, including two-qubit gates activated via fast DC pulses \cite{Arute2020,Dicarlo2019}.
 
The paper is organized as follows. In \cref{sec:SystemHamiltonian}, we describe the system and the corresponding Hamiltonian terms and introduce both the parametric resonance gate and the floating symmetric coupler. \cref{sec:errors} introduces various types of error sources that affect the performance of superconducting qubits and provides analytical expressions needed to estimate their effect on the fidelities of two-qubit gates. We present the experimental results for the CZ gates and analyze its error budget in \cref{sec:experiments}. We finally conclude in \cref{sec:conclusion}. In the Appendix, we provide the derivations for the analytical expressions quantifying different error channels.

\section{Device and Hamiltonian}\label{sec:SystemHamiltonian}
\begin{figure}
    \centering
    \includegraphics{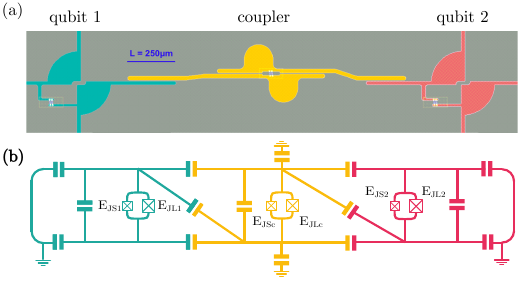}
    \caption{(a) The design layout of two qubits (teal and red) coupled via a floating tunable coupler (yellow)~\cite{Sete_floating_2021}. The gray background is ground, and the teal, red, and yellow shaded regions represent the electrodes of the qubits and the coupler. (b) Simplified circuit representation of the qubit-coupler-qubit system. The floating tunable coupler allows zero-coupling conditions when the coupler frequency is set below that of the qubits, even if the qubit-qubit pitch is approximately 2 mm. This is because the zero-coupling condition for the floating tunable coupler does not depend on the direct qubit-qubit coupling capacitance~\cite{Sete_floating_2021} as opposed to the grounded tunable coupler, where direct qubit-qubit capacitance is required to achieve the zero-coupling condition~\cite{Yan2018}.}
    \label{fig:schematic}
\end{figure}

Our device comprises two tunable transmon qubits interconnected via a symmetric tunable coupler [\cref{fig:schematic}(a)]. The device consists of two separate chips. The qubits, the tunable coupler, and the readout resonators are fabricated on a ``transmon" chip, whereas the microwave control lines, flux control lines,  and readout feedlines are fabricated on a ``control'' chip. These two chips are bonded via indium bump bonds, and the signals to the transmon chip are delivered through signal vias connecting the two chips. Each qubit is individually connected to a dedicated readout resonator, which, in turn, is coupled to a common feedline on the control chip. The readout tone is delivered through indium bumps connecting the two chips. Through silicon vias are used to deliver signals from one side of the control chip to the other. The microwave and flux bias signals are combined using cryogenic diplexers before reaching the chip~\cite{ric2021}. The device can be represented by a simplified circuit,  as depicted in \cref{fig:schematic}(b). The corresponding Hamiltonian is given by~\cite{Sete_para_2021}

\begin{align}
\label{Ham-quant}
H &= \sum_{j=1,2,c}4E_{Cj}\hat n_j^2-E_{Jj}\cos(\hat \phi_{j}+\phi_{0j})\notag\\
&+4E_{1c}\hat n_1 \hat n_c+4E_{2c}\hat n_2 \hat n_c+4E_{12}\hat n_1\hat n_2,
\end{align}
where $\hat n_{j}$ represents the number operator for the Cooper pairs transferred between the superconducting islands of the Josephson junction, and $\hat{\phi}_j$ represents the gauge invariant phase difference across the Josephson junction for the $j$'th element. Similarly, $E_{Cj}$ and $E_{Jj}$ are the charging and the effective Josephson energies, respectively, which are given by
\bes
\begin{align}
\label{EJs}
    E_{Jj}& = \sqrt{E_{JSj}^2+E_{JLj}^2+2E_{JSj}E_{JLj}\cos(\phi_{ej})},\\
    \phi_{0j} &= \tan^{-1}\left[\frac{E_{JSj}-E_{JLj}}{E_{JSj}+E_{JLj}} \tan(\phi_{ej}/2)\right]\label{phi0}, 
\end{align}
\ees
where $j \in \{1,2,c\}$ with $E_{JLj}$ and $E_{JSj}$ being the Josephson energies for the large and small junctions of the SQUIDs. $E_{1c}$, $E_{2c}$ and $E_{12}$ are coupling energies. We also introduce the external phase $\phi_{ej}$, which is related to the external flux $\Phi_{ej}$ through the expression $\phi_{ej} = 2\pi \Phi_{ej}/\Phi_0$ for each qubit and coupler. 

%================ Figure 1 ==========
%===================================

Truncating the Hilbert space to three levels in each of the qubits and the coupler, the Hamiltonian [\cref{Ham-quant}] of the transmon-coupler-transmon system in the harmonic oscillator basis can be reduced to 
\begin{align}
    H/\hbar &= \sum_{j=1,2,c}\omega_{j}|1\rangle_j\langle 1|+ (2\omega_j+\eta_j)|2\rangle_j\langle 2|\notag\\
    &
    + g_{1c}X_{1}X_{c} +g_{2c}X_{2}X_{c} +g_{12}X_{1}X_{2},
\end{align}
where $\omega_j$ are the qubit and coupler frequencies, $\eta_{j}$ are the anharmonicities,  $X_{j} = a_{j}+a_{j}^{\dag}$ with $a_{j} = |0\rangle\langle 1| + \sqrt{2} |1\rangle\langle 2|$. $g_{1c}$ and $g_{2c}$ are qubit-coupler couplings, and $g_{12}$ is a direct qubit-qubit coupling (see \cite[App.~A]{Sete_para_2021}). 

The coupler pads are symmetrically connected to both qubits in a floating tunable coupler architecture with a symmetric configuration. The zero-coupling condition is achieved by placing the coupler frequency below that of the qubits~\cite{Sete_floating_2021}. A parametric resonance gate is activated by applying a modulated flux pulse on one of the qubits and bringing the target transitions into resonance during modulation. For example, a CZ gate is implemented by bringing $|11\rangle$ and $|20\rangle$ or $|02\rangle$ into resonance during modulation. In this work, we focus on a parametric resonance CZ gate. 

%=================== Figure 2 ==================
\begin{figure}
    \includegraphics[width=\linewidth]{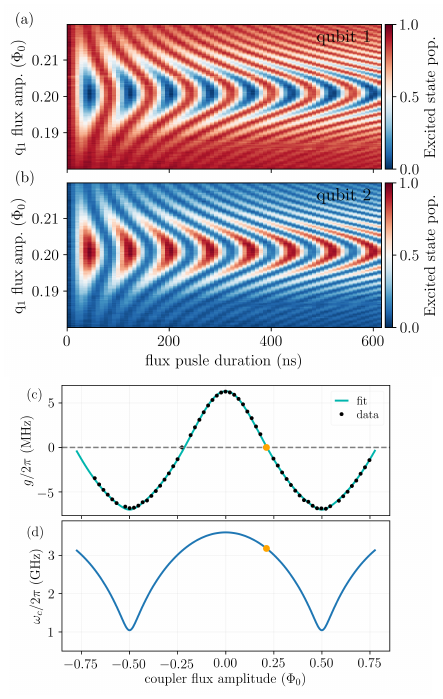}
    \caption{Measured populations of $|10\rangle$ ($|q_{1}q_{2}\rangle$) (a) and of $|01\rangle$ (b) with initial state $|\psi_{0}\rangle=|10\rangle$ as a function of flux pulse duration and flux pulse amplitude of qubit 1 for a fixed coupler flux amplitude. By fitting the oscillations, we extract the coupling strength between the qubits. (c) Net qubit-qubit coupling $g$ versus tunable coupler flux amplitude. The solid teal line is fit to \cref{gqq}, and the black dots represent the experimental data.  (d) Tunable coupler frequency versus fast flux amplitude through SQUID loop. The orange dots show the coupler bias at which zero coupling occurs ($\Phi_{ec} = 0.212~\Phi_0$) and the corresponding coupler frequency  $\omega_{c}/2\pi = 3.18 \mathrm{GHz}$. The chevron shown in (a) is measured at the coupler's maximum frequency.  The parameters extracted from the fits are shown in \cref{tab:qubit-characteristics}.}
    \label{fig:coupling-data}
\end{figure}
%===============================================

%==================== Table 1 ==========
\renewcommand{\arraystretch}{1.5}
\begin{table*}
 \caption{Hamiltonian parameters of the device. The qubit and resonator frequencies are measured using spectroscopy, while the coupler frequencies are extracted using the qubit-qubit coupling versus flux bias fits in \cref{fig:coupling-data}. The coherence times are measured at the maximum frequencies of the transmon qubits (flux sweet spots). Here $T_{2R}$ and 
 $T_{2E}$ are the decoherence times from Ramsey and  Han-echo experiments, respectively. The coupling rates are also extracted from the coupling versus coupler flux bias fit, with the error bars being $1\sigma$ standard deviations.} 
\centering
\begin{tabularx}{\textwidth}{XXXXXXXX}
\hline
\hline
& $\omega_{01}^{\rm max}/2\pi$  & $\omega_{01}^{\rm min}/2\pi$ & $\eta/2\pi$& $\omega_r/2\pi$ & $T_1(\mu \mathrm{s})$ & $T_{2R}(\mu \mathrm{s})$ & $T_{2E}(\mu \mathrm{s})$ \\
&  (MHz) &  (MHz) & (MHz)& (MHz)&  &  &  \\
\hline
qubit 1 & 4576 & 3989 & -203 &7408 & $22 \pm \ 1.9$ & $13 \pm \ 1.1$ & $22 \pm \ 1.8$\\
qubit 2 & 4415 & 3773 & -203 & 7359& $25 \pm \ 1.2$ & $22 \pm \ 1.3$ & $32 \pm \ 1.5$\\
coupler & 3597 & 1044 & -130 & - &- & -& -\\
\hline
\end{tabularx}

\begin{tabularx}{\textwidth}{XXX}
  \centering
   & $g_{12}/2\pi$ (MHz) & $\sqrt{|g_{1c}g_{2c}|}/2\pi$ (MHz) \\
    \hline
Coupling strengths  &$-7.45~\pm~0.13$ & ~~$104.55~\pm~0.36$ \\
    \hline
    \hline
  \end{tabularx}
\label{tab:qubit-characteristics}
\end{table*}

Before delving into characterizing a two-qubit gate, we first measure the zero-coupling condition. To achieve this, we begin by exciting one of the qubits, such that $|\psi_{0}\rangle = |10\rangle$. A modulated flux pulse is applied on the higher frequency qubit to bring it into resonance with the unmodulated qubit, while a fast DC pulse is applied to the tunable coupler to turn on the qubit-qubit coupling. This leads to a population exchange between the initial state $|10\rangle$ and the state $|01\rangle$ as illustrated in \cref{fig:coupling-data}~(a) $\&$ (b). The rate of oscillations at the center of the chevron yields the net coupling between the qubits. By repeating this measurement for a varying amplitude of the coupler flux pulse, we construct a profile of the qubit-qubit coupling as a function of the coupler flux pulse amplitude. The coupler is then parked at a DC flux bias (which is the same as the flux pulse amplitude that gives zero coupling) where the coupling is zero or minimal. This is possible because both the slow and fast flux pulses are generated by the same source. 

The coupling parameters and tunable coupler frequency can be extracted by fitting the qubit-qubit coupling data versus coupler bias to the expression \cite{Sete_para_2021, Sete_floating_2021}
\begin{align}
    g(\phi_{ec}) = g_{12}- \frac{1}{2}g_{1c}(\phi_{ec})g_{2c}(\phi_{ec})\sum_{k=1}^2\left(\frac{1}{\Delta_k} + \frac{1}{\Sigma_k}\right),\label{gqq}
\end{align}
where $\Delta_k = \omega_c (\phi_{ec})-\omega_k$ and $\Sigma_k = \omega_{c}(\phi_{ec})+\omega_k$, and the coupler frequency is $\hbar\omega_c(\phi_{ec}) = \sqrt{8E_{Jc}(\phi_{ec})E_{Cc}}-E_{Cc}(1 + \xi/4)$, where $E_{Jc} = [E_{JSc}^2+E_{JLc}^2+2E_{JSc}E_{JLc}\cos(\phi_{ec})]^{1/2}$ and $\xi =\sqrt{2E_{Cc}/E_{Jc}}$  (see~\cite[Eq.~(B8)]{Sete_para_2021} with $n=0$, the condition for a parametric resonance gate). 

Figure~\ref{fig:coupling-data}(c) shows the experimental data of net qubit-qubit coupling as a function of the amplitude of the coupler flux pulse, clearly identifying the zero coupling condition between the qubits at $\Phi_{ec} = 0.212~\Phi_{0}$ corresponding to $\omega_{c}/2\pi=3.18~\mathrm{GHz}$ obtained from the fit shown in \cref{fig:coupling-data}(d). It should also be noted that the maximum coupling occurs when the coupler is at its minimum frequency, with the coupling governed primarily by $g_{12}$, which follows from \cref{gqq}. We maintain the coupler at zero-coupling bias during idling, where no two-qubit gate operation occurs.

The qubit and coupler parameters and qubit coherence times are summarized in \cref{tab:qubit-characteristics}. Note that the polarities of the couplings are assigned based on knowledge of the design~\cite{Sete_floating_2021} (in a symmetric coupler design, all couplings have a negative sign). 

%===================== Figure 3 ==========================
\begin{figure}
    \centering
    \includegraphics[width=\linewidth]{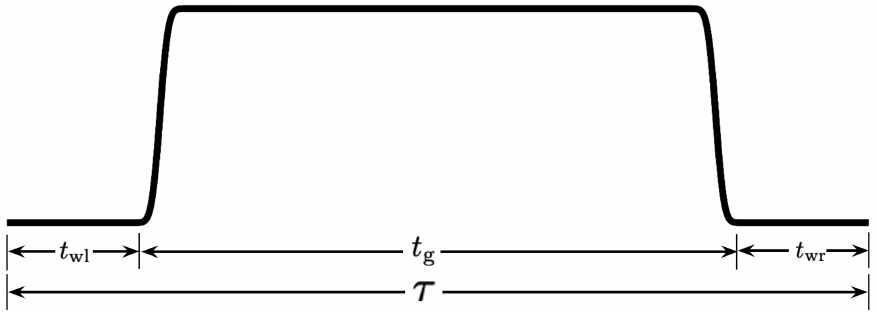}
    \caption{The $\mathrm{erf}$ function flux pulse envelope [\cref{evelope}] applied to the coupler and the qubit showing different parts of the pulse. Wait times before ($t_{\rm wl}$) and after ($t_{\rm wr}$) the pulse are added to avoid RF and flux pulse overlap. The dephasing time of the modulated qubit is different during the wait (idling) and activate parts ($t_{\rm g}$) of the pulse. The total length of the two-qubit gate is defined as $\tau =t_{\rm g}+t_{\rm wl} +t_{\rm wr}$.}
    \label{fig:pulseshape}
\end{figure}
%=========================================================

\section{Sources of Error in parametric resonance gates}
\label{sec:errors}
In this section, we discuss various sources of error that impact the fidelity of the parametric resonance gate, categorizing them broadly into coherent and incoherent errors. It is important to note that the qubits exhibit differing coherence times during idling and gate operation. Therefore, to estimate the magnitude of the incoherent errors, we divide the total gate time, denoted by $\tau = t_{\rm w} + t_{g}$, into idling and active phases, as shown in \cref{fig:pulseshape}. The idle period ($t_{\rm w}=t_{\rm wl} +t_{\rm wr}$) represents the wait time before and after the pulse application, whereas the active phase ($t_{\rm g}$) comprises the rise, fall, and flat part of the pulse. During the gate, the qubit is tuned away from the flux sweet spot to a frequency range more susceptible to flux noise, reducing dephasing times. Here we assume that the qubit coherence is the same during the rise and fall times as well as during the flat part of the flux pulse.

\subsection{Incoherent errors}
\subsubsection{Error due to qubit relaxation: $T_1$ }
Qubit relaxation is a primary error source in superconducting qubits. It can arise from the qubits' coupling to the control lines or from intrinsic material losses. In our device, both the microwave and flux signals are transmitted via a single line on the control chip [40]. Although this line is primarily designed for inductive coupling with the qubit, the presence of a small yet finite capacitive coupling introduces a potential loss channel, affecting the qubit’s relaxation time $T_1$. Additionally, the coupler has a dedicated flux line on the control chip, potentially creating an unintended coupling to the qubit and introducing an extra qubit decay channel. Other contributing factors include the Purcell effect through the readout line. For a CZ gate enacted between states $|11\rangle$ and $|20\rangle$ and for the two-qubit gate time much shorter than the relaxation time ($\tau/T_1 \ll 1$), the error due to qubit relaxation is given by (see \cref{incoh_CZ})
\begin{align}\label{CZT1_error}
    r_{\rm CZ}^{T_{1}} = \frac{2}{5} \left(\frac{1}{T_{1,1}} + \frac{1}{T_{1,2}}\right)t_{\rm w} + \left(\frac{1}{2 \tilde T_{1,1}} +\frac{3}{10\tilde T_{1,2}}\right)t_{\rm g},
\end{align}
where $T_{1,j}$ and $\tilde T_{1,j}$ are the relaxation times of the $j^{\rm th}$ qubit during the idling and active parts of the gate, respectively. The relaxation time of the modulated qubit is measured by applying gate flux pulses on the qubit and the coupler during the delay of the standard $T_1$ measurement protocol, thus simulating the operational conditions of the gate. For a CZ gate activated between states $|11\rangle$ and $|02\rangle$, the corresponding error can be obtained by interchanging the qubit indices in \cref{CZT1_error}.   

Similarly, for an iSWAP gate, the error due to the relaxation of the qubits reads (see \cref{AppendixA})
\begin{align}
    r_{\rm iSWAP}^{T_{1}} = \frac{2}{5}\left(\frac{t_{\rm w}}{T_{1,1}} + \frac{t_{\rm w}}{ T_{1,2}}  \right) + \frac{2}{5}\left(\frac{1}{\tilde T_{1,1}} + \frac{1}{\tilde T_{1,2}}  \right)t_{\rm g} .
\end{align}
We note that when two-qubit gate activation involves non-computational states, such as in the case of the CZ gate, it is important to model transmons as a three-level system when estimating the incoherent error. This is because the excitation might be transferred during the gate to the non-computational states,  which decay or dephase faster. Not including higher transmon states underestimates incoherent errors \cite{Sung2021,Li2024}.

\subsubsection{Error due to qubit dephasing: $T_{\phi}^{\rm wh}$}

In our setup, both the qubits and the coupler are flux-tunable transmon qubits. Flux tunable qubits are known to be susceptible to broadband (white) flux noise, for example, noise originating from the control electronics. Beyond white flux noise, photon number fluctuations within the resonator, known as shot noise, also contribute to qubit dephasing. Here, our focus is specifically on Markovian noise, which manifests itself as an exponential decay of the qubit's superposition state, represented by $\rho_{01} \sim e^{-\Gamma_{2}t}$. Here, $\Gamma_2$ is defined as $\Gamma_2 =\Gamma_{\phi}^{\rm wh} +\Gamma_{1}/2$, where $\Gamma_{1}= 1/T_{1}$ and $\Gamma_{\phi}^{\rm wh} = 1/T_{\phi}^{\rm wh}$. This model provides a framework for understanding and quantifying the impact of environmental noise on the stability of qubit states.

When the qubits are parked at their respective flux sweet spots during idling, they exhibit first-order insensitivity to flux noise. Consequently, during idling periods the shot noise or other flux-independent white noise dictates the dephasing time $T_{\phi, \rm w}$. During the gate operation, one of the qubits' frequencies is tuned from its parking frequency to bring the qubit energy levels into resonance. As a result, the qubit becomes sensitive to flux noise. The degree of this susceptibility is contingent on both the strength of the noise's spectral density and the sensitivity of the qubit's frequency to flux variations ($\partial \omega_{01}/\partial \Phi$). In scenarios where this sensitivity is pronounced, flux noise can become a critical limiting factor for $T_{\phi}^{ \rm wh}$. 

The error in a CZ gate (activated between states $|11\rangle$ and $|20\rangle$) attributable to such dephasing mechanisms, particularly when the gate duration $\tau$ is significantly shorter than $T_{\phi}^{\rm wh}$ ($\tau/T_{\phi}^{ \rm wh}\ll 1$) is quantified as follows (see \cref{incoh_CZ})
\begin{align}\label{CZ20_err}
    r_{\rm CZ}^{T_{\phi}^{\rm wh}} = \frac{2}{5}\left( \frac{1}{T_{\phi,1}^{\rm wh}} +\frac{1}{T_{\phi,2}^{\rm wh}} \right)t_{\rm w}
    + \left(\frac{61}{80 \tilde T_{\phi,1}^{\rm wh}} + \frac{29}{80 \tilde T_{\phi,2}^{\rm wh}}\right) t_{\rm g},
\end{align}
where $T_{\phi,j}$ and $\tilde T_{\phi,j}$ are white noise-induced dephasing time during idle and active parts of the qubit flux pulse, respectively. For a CZ gate operating between states $|11\rangle$ and $|02\rangle$, the error during the gate is obtained by interchanging the indices in \cref{CZ20_err}. Similarly, for an iSWAP gate, the error due to white noise-induced dephasing is given by: 
\begin{align}
    r_{\rm iSWAP}^{T_{\rm \phi}^{\rm wh}} = \frac{2}{5}\left( \frac{t_{\rm w}}{T_{\phi,1}^{\rm wh}} + \frac{t_{\rm w}}{ T_{\phi,2}^{\rm wh}}   \right) +\frac{2}{5}\left( \frac{t_{\rm g}}{\tilde T_{\phi,1}^{\rm wh}} + \frac{t_{\rm g}}{\tilde T_{\phi,2}^{\rm wh}}   \right).
\end{align}

\subsubsection{Error due to $1/\mathrm{f}$ noise dephasing: $T_{\phi}^{1/\rm f}$}

In this section, we consider the error contribution from low-frequency 1/f non-Markovian noise. Historical measurements on SQUIDs~\cite{Koch1983} have revealed the presence of flux noise with a $1/\mathrm{f}$-like noise power spectrum that has been extensively studied since then~\cite{Wellstood87,Devoret2002,Martinis2003,Ithier05,
  Yoshihara_2006,Bialczak2007,RHKoch_2007,Faoro_2008,Manucharyan_2009,
  Martinis_2014}. The noise, characterized by a magnitude of a few $\mu\Phi_0/\sqrt{\rm Hz}$ at $1~\rm{Hz}$, appears to be a universal phenomenon, observed across varied device designs, materials, and sample dimensions~\cite{Wellstood87}. Although the microscopic origin of $1/\mathrm{f}$ flux noise remains elusive, recent theoretical and experimental studies suggest that molecular oxygen on the SQUID surface, acting as free magnetic fluctuators, could be a significant contributor~\cite{Wang15,Kumar16}. Apart from these intrinsic noise sources, low-frequency noise may also originate from control electronics. Because the dephasing rate depends on the sensitivity of the qubit frequency to flux variation. One strategy to mitigate this sensitivity involves designing a tunable transmon with asymmetric SQUID junction energies, as proposed in Ref.~\cite{Plourde17}.

In the limit $\tau/T_{\phi}^{1/\rm f}\ll 1$, the error attributable to $1/\mathrm{f}$ flux noise dephasing for CZ gate activated between $|11\rangle$ and $|20\rangle$ has the form (see Appendix \ref{one_on_f})
\begin{align}\label{CZone_on_f}
    r_{\rm CZ}^{T_{\phi}^{1/ \rm f}} =\frac{61}{80}\left(\frac{t_{\rm g}}{\tilde T_{\phi,1}^{1/\rm f}}\right)^2+ \frac{29}{80}\left(\frac{t_{\rm g}}{\tilde T_{\phi,2}^{1/\rm f}}\right)^2,
\end{align}
where $\tilde T_{\phi,j}^{1/\rm f}$ is the dephasing time due to  $1/\mathrm{f}$ flux noise under modulation. Note that the first term in \cref{CZone_on_f} is the error due to $1/\mathrm{f}$ flux noise for qubit 1, assuming that the qubit is initially parked away from the flux sweet spots. However, if the qubit starts at the sweet spot and does not undergo a qubit frequency shift due to hybridization with the coupler, it remains largely insensitive to flux noise.  For the gate actuated between $|11\rangle$ and $|02\rangle$, the error rate due to $1 \rm f$ can be obtained by exchanging the qubit indices in \cref{CZone_on_f}.

\begin{figure}
    \centering
    \includegraphics{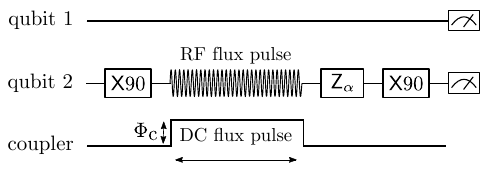}
    \caption{Pulse sequence diagram employed to measure qubit dephasing under flux modulation. The sequence initiates with an X90 pulse, followed by a modulated flux pulse applied to the qubit and a DC flux pulse applied on the coupler with amplitude matching that during gate operation.  Subsequently, a virtual Z rotation is introduced before the second X90 pulse to facilitate the observation of Ramsey oscillations. The resulting measured data is fitted to $\propto \exp[-\Gamma_2 t- (\Gamma_{\phi,1/\rm f}t)^2]\cos(\delta t)$, allowing us to quantify the dephasing rates.}
    \label{fig:Ramsey}
\end{figure}

Similarly, for an iSWAP gate, the error due to $1/\mathrm{f}$ noise dephasing is expressed as
\begin{align}
   r_{\rm iSWAP}^{T_{\phi}^{1/ \rm f}} = \frac{2}{5}\left[\left(\frac{t_{\rm g}}{\tilde T_{\phi,1}^{1/ \rm f}}\right)^2 + \left(\frac{t_{\rm g}}{\tilde T_{\phi,2}^{1/ \rm f}}\right)^2\right].
\end{align}
To quantify the dephasing of the qubit during gate operation, we applied a modulated flux pulse to one of the qubits within the delay period between the two X90 pulses in a Ramsey experiment as depicted in \cref{fig:Ramsey}. To accurately capture the gate conditions, we also apply a flux pulse on the coupler, matching the amplitude used for gate activation (see \cref{fig:Ramsey}). The Ramsey data are then fitted to the model $\rho_{01} \propto \exp[-\Gamma_2 t- (\Gamma_{\phi,1/\rm f} t)^2]\cos(\delta t)$. This approach isolates the dephasing contribution exclusively attributable to $1/\mathrm{f}$ flux noise and white noise.

%======================== Figure 4 ===================
%======================== Figure 4 ===================

\subsection{Coherent errors}
\subsubsection{Amplitude and phase errors}
Amplitude and conditional phase errors represent two prevalent sources of coherent errors for parametric resonance gates. The amplitude errors typically arise from incomplete population transfer between the intended energy levels. In contrast, conditional phase errors often result from residual $ZZ$ interactions, which may lead to either under-rotation or over-rotation for a CZ gate and a nonzero conditional phase for an iSWAP gate. To quantify the amplitude and phase errors, one can compare the target unitary operation with the realized fSim (fermionic simulation) unitary. The amplitude error, attributed to deviations in the rotation angle, is calculated as:
\begin{align}
    r_{\delta \theta} = \frac{2}{5}[3+\cos(\delta \theta)]\sin^2(\delta\theta/2),
\end{align}
where $\delta \theta$ represents the difference between the intended and the actual rotation angles. Similarly, the phase error, which accounts for the discrepancies in the conditional phase shift, is given by~\cite{Sete_para_2021}
\begin{align}
    r_{\delta \phi} = \frac{3}{10} [1-\cos(\delta\phi)].
\end{align}
where $\delta \phi$ denotes the variance between the target and the actual conditional phases. 

%======================== Figure 5 ===================
\begin{figure}
    \centering
    \includegraphics[width=\linewidth]{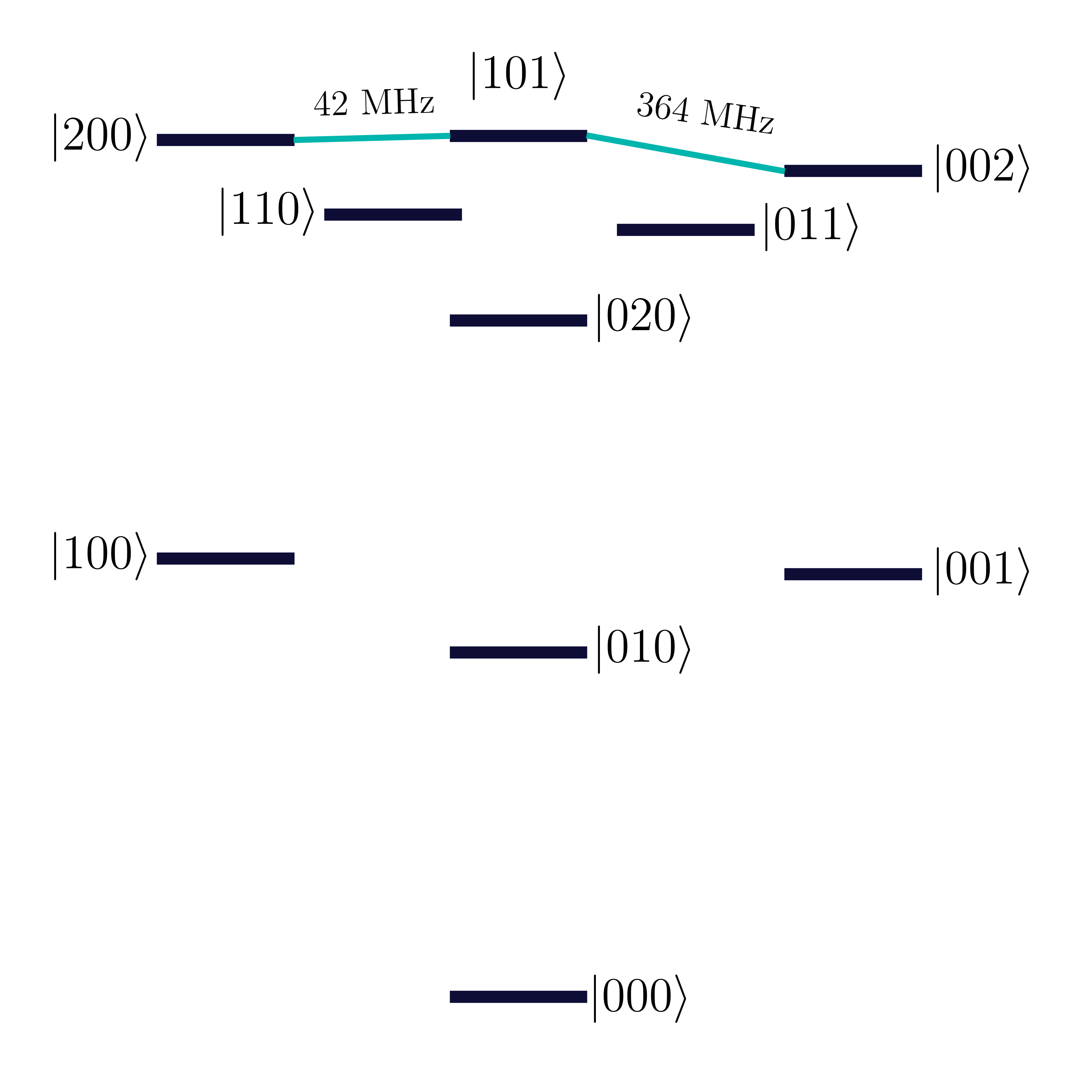}
    \caption{Energy level diagram illustrating the states of the qubits and coupler ($|nlm\rangle = |q_1 q_c q_2\rangle$). CZ gates can be activated by bringing either $|101\rangle$ and $|200\rangle$ or $|101\rangle$ and $|002\rangle$ into resonance for a full swapping period under flux modulation. Note that a CZ gate between $|101\rangle$ and $|200\rangle$ is preferable because of the smaller detuning, which means less deviation of the modulated qubit 2 from the flux sweet spot, and hence less sensitivity to flux noise. An iSWAP gate can be activated by bringing $|100\rangle$ and $|001\rangle$ in resonance for half a swapping period.}
    \label{fig:energy_level}
\end{figure}
%======================== Figure 5 ===================

%======================== Figure 6 ===================
\begin{figure*}
    \centering
    \includegraphics[width=\linewidth]{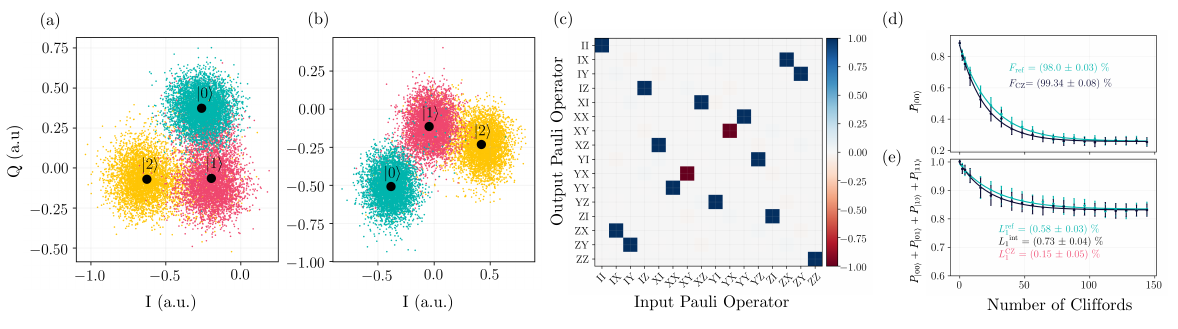}
    \caption{Three-state readout response data for qubit 1 (a) and qubit 2 (b). The three-state classification of the IQ data is used for leakage error characterization. Qubit 1 [2] three-state classification has a fidelity of $92.2 \%$ [$87.5\%$] with a measurement time of $2.7 \mu \mathrm{s}$ [3.5 $\mu\mathrm{s}$]. (c) Estimated quantum process tomography for CZ gate [gate time, $\tau = 64 \mathrm{ns}$] with a measured conditional phase $\phi=3.086$ radians and average quantum process fidelity of $90.73 \%$. (d) Reference randomized benchmarking (RB) and interleaved randomized benchmarking (iRB) with fidelities $F_{\rm ref} = 98.00 \pm 0.03 \%$ and $F_{\rm CZ} = 99.34 \pm 0.08 \%$, respectively. (e) Leakage randomized benchmarking for the same CZ gate, highlighting the leakage error in both leakage RB and leakage iRB experiments, with the net leakage error for the CZ gate being $L_{1}^{\rm CZ}= 0.15\pm 0.05 \%$}.
    \label{fig:three-state-cal}
\end{figure*}
%======================== Figure 6 ===================
\subsubsection{Leakage error}
We investigate the effects of population leakage beyond the computational states of qubits, namely $|00\rangle,|01\rangle,|10\rangle,|11\rangle$. In a parametric resonance gate,  leakage can occur through two primary mechanisms. The first involves unintended resonance crossing during the rise and fall times of the qubit's flux pulse. For example, when a qubit is initially prepared in an excited state and a flux pulse is applied to bring its frequency into resonance with the other qubit, it may traverse undesired resonances during the pulse's rise time, resulting in leakage into nontargeted states.

The second mechanism that causes potential leakage involves interaction between the $|11\rangle$ state and higher energy levels of the qubits and states involving the qubit-coupler. Our setup lacks a dedicated readout resonator for the coupler, precluding precise quantification of leakage into coupler states through straightforward approaches. However, since the coupler typically remains in its ground state and is detuned by more than 1 GHz below the qubit frequencies, leakage to the coupler is expected to be minimal. Our device simulations, based on measured parameters, indicate that the most substantial leakage occurs into the $|02\rangle$ state, closely followed by the next nearest (in terms of energy) state $|20\rangle$ as illustrated in \cref{fig:energy_level}.

To quantify leakage beyond the computational subspace, we employ the leakage randomized benchmarking (RB) protocol~\cite{Cross2018}. Unlike the standard RB measurement~\cite{Magesan2012} that focuses on the survival probability of the ground states at the end of the RB sequence, the leakage RB measures the probability that the qubits stay within the computation subspace, i.e., $P_{\rm cs} =P_{|00\rangle}+P_{|01\rangle} +P_{|10\rangle} +P_{|11\rangle}$. Without leakage, this probability should be close to 1, with deviations primarily due to incoherent errors and readout fidelity. The gate error is deduced from the leakage error rates obtained through RB and interleaved RB (iRB) experiments. By fitting the subspace probability $P_{\rm cs}$ for the RB and iRB experiments to $P_{\rm ref} = b_{\rm ref} + a_{\rm ref}~ p_{\rm ref}^{N_{c}}$ and $P_{\rm int} = b_{\rm int} + a_{\rm int}~ p_{\rm int}^{N_{c}}$, respectively, the corresponding leakage errors are computed as~\cite{Oliver21}
\begin{align}
    L_1^{\rm ref} &= (1-b_{\rm ref}) (1 - p_{\rm ref}),\\
    L_1^{\rm int} &= (1-b_{\rm int}) (1 - p_{\rm int}).
\end{align}
The target two-qubit gate error due to leakage can then be calculated using
\begin{align}
    L_{1}^{\rm gate} = 1 -\frac{1-L_1^{\rm int}}{1-L_1^{\rm ref}}.
\end{align}
To estimate the leakage error from the RB leakage protocol, one needs to use a three-state classification for each qubit. In \cref{fig:three-state-cal}, we show an example of three-state readout response data for each qubit on which we run a three-way linear classification for leakage error analysis.

\section{error budget: Experiment}
\label{sec:experiments}
To demonstrate the error analysis of a parametric resonance gate within a tunable coupler architecture, we specifically focus on studying a CZ gate experimentally. The energy level diagram of the qubit-coupler-qubit system (see \cref{fig:energy_level}) indicates two possible activation mechanisms for a CZ gate: using either $|101\rangle \leftrightarrow |200\rangle$ or $|101\rangle\leftrightarrow |002\rangle$ transitions, corresponding to detunings of $42$ MHz and $364$ MHz, respectively. The preferred option is the CZ gate enacted between the levels with the smaller $42$ MHz detuning. It is preferred because the modulated qubit frequency excursion to achieve the CZ resonance condition is lower than the other CZ gate, thereby exhibiting reduced sensitivity to flux noise. Moreover, the $364$ MHz detuning pathway risks encountering the iSWAP resonance, potentially leading to unwanted leakage transitions. Consequently, our experimental investigations will focus on the CZ gate activated between $|101\rangle$ and $|200\rangle$ states. 

We activate the parametric resonance CZ gate by sending a modulated flux pulse on qubit 2, described as follows:  
\begin{align}
 \Phi_2(t) = 
    \begin{cases}
        & 0, ~~  t < t_{\rm wl}\\
         & \Phi_m f(t) \cos(\omega_m t), ~~   t_{\rm wl} \leq t \leq \tau +t_{\rm wr},\\
        &0, ~~ t> \tau +t_{\rm wr},
    \end{cases} 
\end{align}
where $\Phi_m$ and $\omega_m$ are the flux modulation amplitude and frequency of the flux pulse, respectively. The pulse envelope $f(t)$ is an error function (erf) pulse given by 
\begin{align}\label{evelope}
f(t)= \frac{1}{2} \left[\mathrm{erf}\left(\frac{t-t_1-t_{\rm wl}}{\sigma}\right) - \mathrm{erf}\left(\frac{t-t_2 -t_{\rm wr}}{\sigma}\right)\right],
\end{align}
where $t_{1} = t_{r}/2$, $t_{2} = t_{\rm f} - t_{1}$, and $\sigma=t_{1}/[2\sqrt{2 \ln(2})]$. Here, $t_{r}= 4~\mathrm{ns}$ is the rise time of the pulse. A fast DC flux pulse is also applied to the coupler $\Phi_{c}(t) = \Phi_{c} f(t)$ with $\Phi_{c}$ being its amplitude, as needed to turn on the qubit-qubit coupling to achieve the desired gate time. Since the parametric resonance gate does not depend on the flux pulse modulation frequency~\cite{Sete_para_2021}, we choose $\omega_{m}/2\pi = 250~\mathrm{MHz}$ to avoid sideband transitions. The modulation frequency will be fixed for all gates characterized in this work.

\begin{table*}[]
\caption{This table summarizes the coherence times of qubits 1 and 2 during both idling and active gate operation periods. For qubit 1, the coherence times are consistent across both idling and during gate, while for qubit 2, they vary depending on whether it is the idling period or gate period. The $1\sigma$ error bars represent the standard deviation of repeated qubit coherence measurements. Note that qubit 1 is parked at the flux sweet spot and thus is first-order insensitive to $1/\rm f$ flux noise.  }

    \centering
\begin{tabularx}{\textwidth}{X|X}
\hline
\hline
      During idling ($t_{\rm w}$)   & 
      During gate ($t_{\rm  g}$) \\ 
      \hline
\end{tabularx}
\begin{tabularx}{\textwidth}{XXX|XXX}
    &$T_{1}(\mu \mathrm{s})$ & $T_{2R} (\mu \mathrm{s})$ & $\tilde T_{1}(\mu \mathrm{s})$  &  $\tilde T_{2R} (\mu \mathrm{s})$&  $\tilde T_{\phi, 1/\rm f} (\mu \mathrm{s})$ \\
    \hline
qubit 1     & $23.9 \pm 5.3$ & $13.1 \pm  2.8$  & $23.9 \pm  5.3$  & $13.1 \pm 2.8$   & -\\
qubit 2     & $23.0 \pm 1.5$ & $20.0 \pm  0.6$  & $23.4 \pm  2.9$  & $18.8 \pm 2.3$   & $28 \pm 4.8$ \\
\hline
\end{tabularx}
    \label{tab:coherence}
\end{table*}

Let us consider a specific example of a CZ gate with gate time $\tau = 64~ \rm {ns}$, which includes a total of $16~{\rm ns}$ of waiting, or padding before and after the flux pulse. To achieve this gate duration, a coupler flux pulse with amplitude $\Phi_{c} =-0.212~ \Phi_{0}$ and envelope $f(t)$, is applied to tune the coupler frequency to $\omega_{c}/2\pi = 3.597~\mathrm{GHz}$ from its parking frequency of 3.18 GHz [See \cref{fig:coupling-data}]. The local phases are measured using quantum process tomography [\cref{fig:three-state-cal}(c)] and the RF pulses of the qubits are updated with virtual Z rotation to undo these phase accumulations. The measured conditional phase and swap angle from quantum process tomography are $\phi = 3.086$ rad and $\theta=-0.015$ rad with an average process fidelity of $90.73\%$. The phase error due to under rotation $\delta \phi = \pi -3.086 = 0.056 $ rad results in $r_{\delta \phi} = 0.05 \%$ and due to swap angle error $r_{\delta \theta} = 0.01 \%$, indicating that the gate is well calibrated and has low residual $ZZ$ crosstalk interactions. 

The leakage  RB measurements, as shown in  \cref{fig:three-state-cal}(e) yield a leakage error rate of $L_{1}^{\rm CZ} = 0.15 \pm 0.05 \%$. Consequently,  the total coherent error is primarily due to the leakage, given by

\begin{align}
    r_{\rm coh.} &= r_{\delta \theta} + r_{\delta \phi} +L_{1}^{\rm CZ} = 0.21 \pm 0.05 \% . 
\end{align}
The CZ gate, benchmarked via interleaved randomized benchmarking  [\cref{fig:three-state-cal}(d)], has a maximum RB fidelity of $F_{\rm iRB} = 99.34\pm 0.08 \%$. To capture the run-to-run variations, we repeated the iRB measurements that yielded an average CZ error of $r_{\rm iRB} = 0.71 \pm 0.08 \%$ over four different measurements. The relaxation times of the qubits during idle and gate operation are provided in \cref{tab:coherence}. Using these values with \cref{CZT1_error}, given $t_{\rm g} = 48 ~\mathrm{ns}$ and $t_{\rm w} = 16 ~\mathrm{ns}$, CZ error due to qubit relaxation is calculated to be $ r_{\rm CZ}^{T_{1}} = 0.19 \pm 0.06 \%$.

The contribution of the dephasing processes is calculated using the measured dephasing times reported in \cref{tab:coherence} and \cref{CZ20_err} is $r_{\rm CZ}^{T_{\phi}^{\rm wh}}=0.29 \pm 0.07 \%$. Finally, since only the second qubit is modulated and sensitive to $1/\mathrm{f}$ noise, the dephasing error due to $1/\mathrm{f}$ flux noise is $r_{\rm CZ}^{T_{\phi,1/\rm f}} = 0.002 \pm 0.0001 \%.$ Thus, the total incoherent error of the CZ gate is
\begin{align}
    r_{\rm incoh} = r_{\rm CZ}^{T_{1}} + r_{\rm CZ}^{T_{\phi}^{\rm wh}} + r_{\rm CZ}^{T_{\phi,1/\rm f}}  = 0.48 \pm 0.08\%.
\end{align}
As a result, the total estimated error for the gate is $r_{\rm coh} +r_{\rm incoh} = 0.69 \pm 0.09\%$, aligning closely with the measured error of $r_{\rm iRB} = 0.71 \pm 0.08 \%$.

%======================== Figure 8 ===================
\begin{figure}
    \centering
    \includegraphics[width=\linewidth]{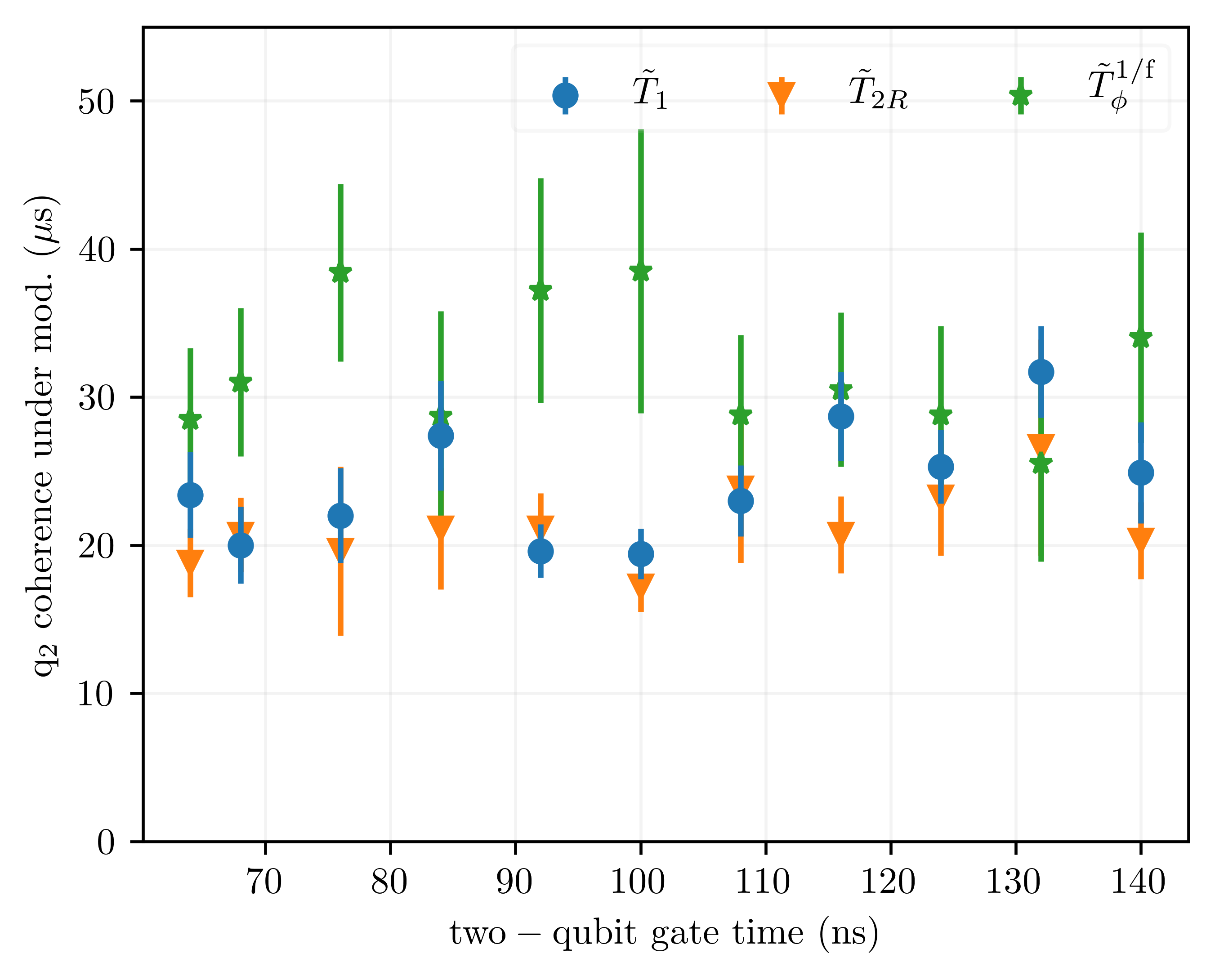}
    \caption{Qubit 2 decoherence times under flux modulation as a function of the two-qubit gate time. The plot highlights how different decoherence times are affected by the length of the gate operation, providing insights into the interplay between gate duration and qubit stability under flux modulation. Note that each gate time corresponds to a different flux pulse amplitude applied to the tunable coupler and thus different coupler frequencies. Because of the coupler-qubit hybridization, the qubit frequency shifts by a few MHz depending on the coupler frequency, leading to qubit coherence variations. }
    \label{fig:coherence}
\end{figure}
%======================== Figure 8 ===================

%======================== Figure 9 ===================
\begin{figure}
    \includegraphics[width=\linewidth]{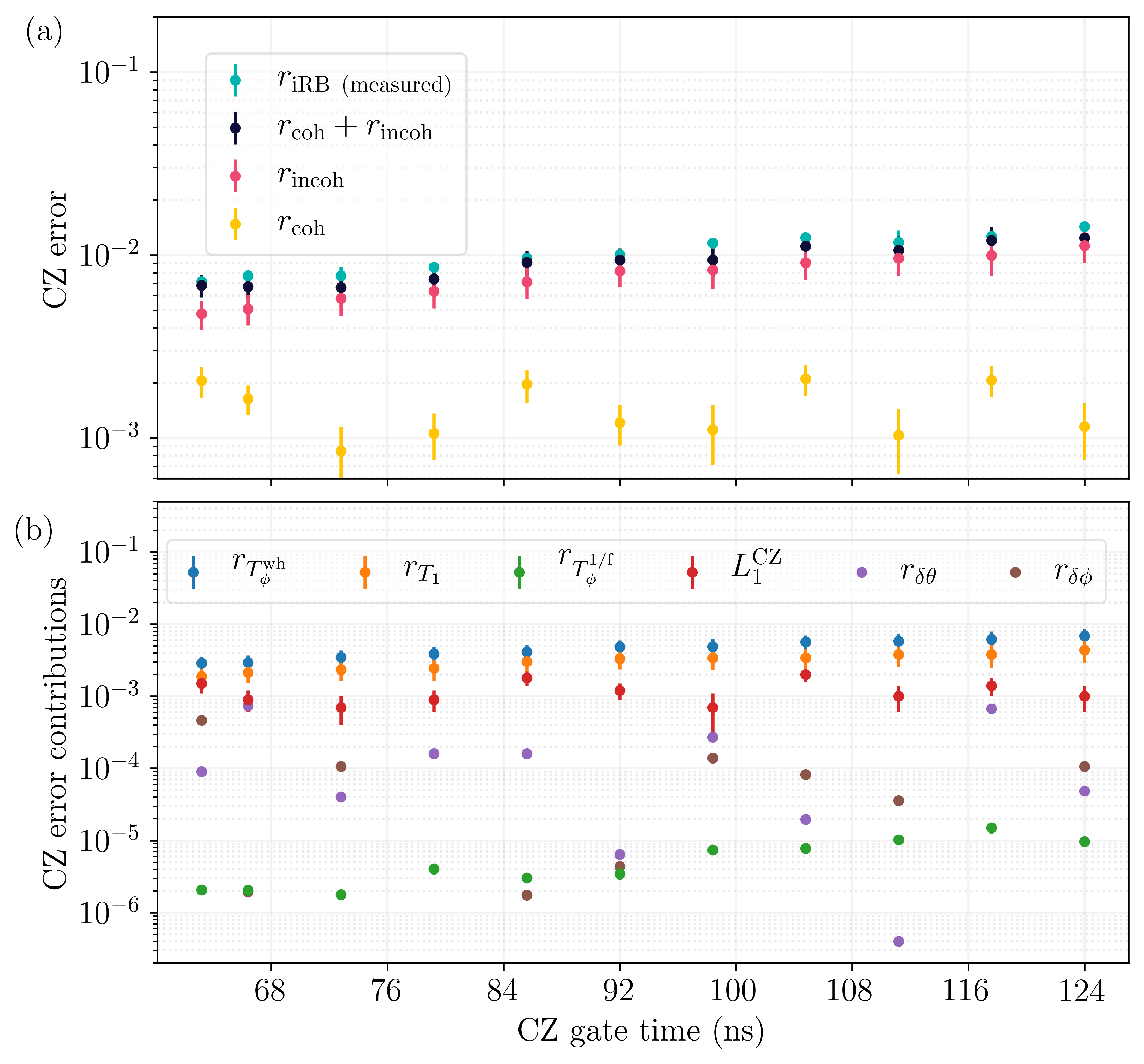}
    \caption{(a) Measured average error rates for CZ gates from interleaved RB ($r_{\rm iRB}$) (teal dots) for various two-qubit gate durations. The plot clearly shows that incoherent errors (red) constitute the majority of the gate error, approximately $83\%$, while coherent errors (yellow) account for the remainder $17\%$. The black dots represent the total estimated error ($r_{\rm coh} +r_{\rm incoh}$), which aligns with the experimentally measured CZ error (teal). (b)  Detailed breakdown of the estimated incoherent and coherent error contributions for each gate time. In particular, $T_1$ (relaxation) and $T_{\phi}^{\rm wh}$ (dephasing due to white noise)  are the dominant error sources.}
    \label{fig:infidelity}
\end{figure}
%======================== Figure 9 ===================

To thoroughly understand the dynamics of various error sources affecting our CZ gate, we conducted a benchmarking study by varying the gate times and performing a detailed error analysis for each duration.  This is achieved by progressively tuning the coupler frequency towards its values corresponding to the zero-coupling condition. As depicted in \cref{fig:coupling-data}(c), the qubit-qubit coupling becomes weaker as the coupler frequency approaches the zero-coupling condition, thus allowing for longer gate times. Additionally, we measured the coherence times of the modulated qubit (qubit 2) under operating conditions for different gate times, as shown in \cref{fig:coherence}.  Although the $\tilde T_{1}$'s do not have a strong dependence on the gate time (or coupler flux amplitude), for gate times $\leq 100$ ns $\tilde T_{1}$ is slightly shorter than for the longer gates. This may be due to the coupler loss impacting the qubit coherence as a Purcell-like effect as the coupler is pulsed towards its maximum frequency which is closer to the qubits. For a few data points ($\tau$ = 92, and 100 ns), we measured consistently lower $\tilde T_{1}$, which could be due to a TLS interacting with the coupler. Note that when parametric gates \cite{Sete_para_2021} (activated with modulated flux pulse) and gates activated by base-band bipolar flux pulses \cite{Dicarlo2019} are operated on qubits parked at the sweet spot, the flux pulse has a refocusing effect, filtering low-frequency dc $1/\rm f$ flux noise \cite{Dicarlo2019, Didier2019}. Our dephasing under modulation measurement captures this echo effect. There could still be low-frequency amplitude noise from the flux pulse itself. We also note that randomized benchmarking itself has a refocusing effect on low-frequency noise and suppresses non-Markovian 1/f noise, whose contribution to the overall infidelity is negligible and hence does not affect our results.

Applying the methodology described earlier for the gate time of $\tau = 64~ \mathrm{ns}$ and using the measured coherence times (\cref{tab:coherence} and \cref{fig:coherence}), we meticulously characterize the error budget for each CZ gate [see \cref{fig:infidelity}]. As illustrated in \cref{fig:infidelity}(a), the experimentally measured average CZ errors (teal dots) align closely with our estimated total error (the sum of coherent and incoherent errors, $r_{\rm coh} +r_{\rm incoh}$). The relative error between the mean of the measured error rate $r_{\rm iRB}$ by RB and the mean of the total estimated error $r_{\rm coh}+r_{\rm incoh}$ defined by $|\mathrm{mean}(r_{\rm iRB}) -\mathrm{mean}(r_{\rm coh}+r_{\rm incoh})|/\mathrm{mean}(r_{\rm iRB})$ yields $10\%$. The incoherent errors (red dots) constitute approximately $83\%$ of the total CZ error on average, and coherent errors account for the remainder $17\%$. In \cref{fig:infidelity}(b), we present a detailed breakdown of each error source that contributes to the CZ gate errors. The major contributions come from qubit relaxation $T_{1}$, dephasing $T_{\phi}^{\rm wh}$ due to Markovian noise, and leakage errors.  The impact of amplitude, phase, and dephasing due to $1/\mathrm{f}$ flux noise remains below $2 \%$ of the total error. 

This comprehensive analysis provides valuable insights into the error dynamics of the CZ gates based on parametric resonance interaction, highlighting the significant role of incoherent errors and leakage, and underscoring the need for optimized control strategies to mitigate these errors. 

\section{conclusion}
\label{sec:conclusion} 
In this work, we have conducted an in-depth experimental analysis of the error budget associated with parametric resonance gates, specifically focusing on flux-tunable qubits in a tunable coupler architecture. We developed a systematic methodology to meticulously quantify various error sources, thereby providing a holistic view of the factors influencing the fidelity of the two-qubit gate. Our investigation is particularly focused on the CZ gate, examining how its performance varies with gate duration.

The findings of our study indicate that incoherent errors, mainly arising from qubit relaxation and dephasing induced by white noise, are predominant factors affecting the fidelity of the CZ gate, accounting for $83\%$ of the gate error. Coherent errors, especially those arising from leakage to noncomputational energy states of the qubits, contribute to the total error rate, accounting for about $17\%$. The experimentally measured average error rates align closely with the total estimated error obtained by analyzing both incoherent and coherent error channels. 

Furthermore, we present practical analytical expressions to estimate both incoherent and coherent errors. These expressions can be readily applied using easily accessible experimental data such as coherence times, swap angles, and conditional phases. This enables a straightforward estimation of two-qubit gate errors using standard measurement protocols, enhancing our ability to predict and mitigate errors in quantum gate operations.

Overall, our work contributes to a better understanding of the error mechanisms in two-qubit gates and provides valuable insights for the optimization and enhancement of quantum gate fidelity.

\begin{acknowledgements}
We thank Nicolas Didier for useful discussions on the derivation of incoherent error expressions.  We also thank Stefano Poletto for critical reading of the manuscript, and Riccardo Manenti and Shobhan Kulshreshtha for their useful discussions on three-state readout classifications. The views, opinions, and/or findings expressed are those of the author(s) and should not be interpreted as representing the official views or policies of the Department of Defense or the U.S. Government. This research was developed with funding from the Defense Advanced Research Projects Agency under Agreement HR00112230006.
\end{acknowledgements}

\appendix

\section{Incoherent error for an iSWAP gate}
\label{AppendixA}

First, we summarize a few pertinent facts about vectorization and the process fidelity. All relevant background can be found, e.g., in Ref.~\cite{wood2015tensor}.

Consider the Lindblad master equation written as 
\begin{align}
\label{eq:Lind}
    \dot \rho(t) = -i[H,\rho(t)] &+ \frac{1}{2}\sum_{k}\gamma_{k}\mc{D}[L_{k}]\rho(t)\notag \\&+\sum_{k}\gamma_{\phi,k}\mc{D}[L_{k}]\rho(t)  ,
\end{align}
where the terms in the Lindblad dissipator are 
\begin{align}
\label{eq:Lind-diss}
\mc{D}[L]\cdot = 2 L \cdot L-\{L^{\dag}L, \cdot\} .
\end{align}

Recall Roth's Lemma~\cite{Roth:1934aa}, which is used to column-vectorize a matrix product of matrices $A$, $B$, and $C$ of compatible dimensions $k\times l$, $l\times m$, and $m\times n$ written in an orthonormal basis:
\begin{align}
\label{eq:Roth}
\overrightarrow{A B C}= (C^T\otimes A) \vec{B} .
\end{align}
Replacing $B$ by $\rho$, the master equation can then be written in column vector form as 
\begin{align}
\label{master_vec_2}
    \dot{\vec{\rho}} = \mc{L}\vec{\rho},
\end{align}
where the matrix $\mc{L}$ is given by
\begin{equation}
\begin{aligned}
\label{eq:L-mat}
    \mc{L} = & -i \mathbbm{1}\otimes H + i H^{T}\otimes\mathbbm{1}\\
    &+\frac{1}{2}\sum_{k}\gamma_{k}(2(L_{k}^{\dag })^T\otimes L_{k}-\mathbbm{1}\otimes  L_k^{\dag}L_k- L_k^{\dag}L_k\otimes \mathbbm{1}) \\
    &+\sum_{k}\gamma_{\phi,k}(2(L_{k}^{\dag })^T\otimes L_{k}-\mathbbm{1}\otimes  L_k^{\dag}L_k- L_k^{\dag}L_k\otimes \mathbbm{1}) .
\end{aligned}
\end{equation}\
To see this, note that using \cref{eq:Roth}, e.g., $\rho H = I \rho H \mapsto (H^{T} \otimes  \mathbbm{1})\vec{\rho}$ under vectorization.  The other terms in \cref{eq:L-mat} follow similarly.

Since the Lindblad equation generates a CPTP map $\mc{E}$, the solution of \cref{eq:Lind} can be written in the Kraus operator sum representation form as
\begin{align}
\rho(t) = \mc{E}[\rho(0)] = \sum_i K_i \rho(0) K_i^\dag ,
\end{align}
where the Kraus operators $\{K_i\}$ satisfy $\sum_i K_i^\dag K_i = \mathbbm{1}$. Vectorizing using \cref{eq:Roth}, this yields
\begin{align}
\overrightarrow{\rho(t)} = \overrightarrow{\mc{E}[\rho(0)]} = \sum_i K_i^* \otimes K_i \overrightarrow{\rho(0)} ,
\end{align}
while the solution of \cref{master_vec_2} has the form
\begin{align}
\overrightarrow{\rho(t)} = e^{\mc{L}t}\overrightarrow{\rho(0)} = \mc{S}\overrightarrow{\rho(0)},
\end{align}
where $\mc{S} \equiv e^{\mc{L}t}$ is the Liouville superoperator matrix. Comparing, we obtain $\mc{S} = \sum_i K_i^* \otimes K_i$.

The average gate fidelity of a gate $U$ relative to a CPTP map (or quantum channel) $\mc{E}$ is defined as~\cite{Nielsen2002} 
\begin{equation}
\begin{aligned}
\bar{F}_{\mc{E},\mc{U}} &= \int d\psi\bra{\psi}U^\dag \mc{E} (\ketb{\psi}{\psi})U\ket{\psi}  \\
&= \int d\psi\bra{\psi}\mc{U}^\dag \mc{E} (\ketb{\psi}{\psi})\ket{\psi} = \bar{F}_{\mc{E}',\mc{I}} \equiv \bar{F}_{\mc{E}'},
\end{aligned}
\end{equation}
where $\mc{U}(\rho) = U\rho U^\dag$, $\mc{E}' \equiv \mc{U}^\dag \mc{E}$, $\mc{I}(\cdot) = \mathbbm{1}\cdot\mathbbm{1}$ is the identity channel, and the integral is over the uniform (Haar) measure $d\psi$, normalized so that $\int d\psi =1$. Thus, $\bar{F}_{\mc{E}'}$ compares $\mc{U}^\dag\mc{E}$ with the identity channel. Note that $\sigma = U^\dag \rho U$ becomes $\vec{\sigma} = (U^T\otimes U^*)\vec{\rho}$ after vectorization. Therefore
\begin{equation}
\begin{aligned}
\overrightarrow{\mc{E}'[\rho(0)]} &= \overrightarrow{U^\dag \mc{E}[\rho(0)]U} = (U^T\otimes U^*)\overrightarrow{\mc{E}[\rho(0)]} \\
&= u^* \mc{S} \overrightarrow{\rho(0)} =\mc{S}'\overrightarrow{\rho(0)} ,
\end{aligned}
\end{equation}
where $\mc{S}' \equiv u^* \mc{S}$ and $u \equiv U^{\dag}\otimes U$ is the Liouville superoperator matrix of the ideal gate $U$.

$\bar{F}_{\mc{E}'}$ can be expressed in terms of $\mc{S}'$ as~\cite{Emerson:2005sf} (see also Ref.~\cite[Eq.~(5.57)]{wood2015tensor})
\begin{align}\label{fidelity}
\bar{F}_{\mc{S}'} = \frac{\Tr(\mc{S}')+d}{d(d+1)} ,
\end{align}
where $d$ is the system's Hilbert space dimension.

Assuming that there are no coherent errors due to the $XY$ interaction, we can estimate the iSWAP fidelity limited by the qubits' relaxation and dephasing rates. We assume that all the relaxation and dephasing rate values are at gate operating conditions. With relaxation and dephasing (Markovian noise), the time evolution of the system is described by the master equation
\begin{align}
\label{master_eqn_1}
    \dot \rho(t) &= -i[H_{\rm iSWAP},\rho(t)] + \frac{1}{2}\sum_{k=1}^2\Gamma_{1,k}\mc{D}[a_{k}]\rho(t) \notag\\
    &+\sum_{k=1}^2\Gamma_{\phi,k}\mc{D}[n_{k}]\rho(t) ,
\end{align}
where $H_{\rm iSWAP} = g(|10\rangle\langle 01| +|01\rangle\langle 10|)$, $n_{k}=a_{k}^{\dag}a_{k}$ ($a_{k} =(|0\rangle\langle 1|)_k$ and $n_{k} = (|1\rangle\langle 1|)_{k}$), and $\Gamma_{1,k}$  and $\Gamma_{\phi,k}$ are relaxation and dephasing rates of the $k_{\rm th}$ qubit, respectively. 

We are interested in both the coherent and incoherent contributions to $\bar{F}_{\mc{E},\mc{U}_{\rm iSWAP}}$, the average gate fidelity of the iSWAP gate, hence now set 
$u = U_{\rm iSWAP}^{\dag}\otimes U_{\rm iSWAP}$.

Since the two-qubit gate times are shorter than the coherence times of the qubits, $t_{\rm g} =\pi/(2g) \ll 1/\Gamma_{1},\ 1/\Gamma_{\phi}^{ \rm wh} $, we can expand the fidelity in small parameters, $\Gamma_{1,k}t_{\rm g}$ and $\Gamma_{\phi,k}^{\rm wh}t_{\rm g}$. In this limit, the loss and dephasing processes are independent of each other in the leading order. Thus, the total super-operator matrix can be written as the sum of the  individual contributions:  
\begin{equation}
\mc{S}_{\rm iSWAP} \approx \sum_{k}\left(\mc{S}_{\mc{L}[\sigma_{k}]} + \mc{S}_{\mc{L}[n_{k}]}\right). 
\end{equation}
We first find the superoperator matrix $\mathcal{S}$ by exponentiating the superoperator $\mc{L}$ given by \cref{eq:L-mat}: 
\begin{align}
    \mc{L} = & -i \mathbbm{1}\otimes H_{\rm iSWAP} + i H_{\rm iSWAP}^{T}\otimes\mathbbm{1}\notag\\
    &+\frac{1}{2}\sum_{k=1}^2\Gamma_{1,k}(2a_{k}\otimes a_{k}-\mathbbm{1}\otimes  n_{k}- n_{k}\otimes \mathbbm{1}) \\
    &+\sum_{k=1}^2\Gamma_{\phi,k}(2 n_{k}\otimes  n_{k} - \mathbbm{1}\otimes  n_{k}^2- n_{k}^2\otimes \mathbbm{1}),\notag
\end{align}
and
\begin{align}
\label{SiSWAP}
    \mc{S}_{\rm iSWAP} = e^{\mc{L}t_{\rm g}}.
\end{align}
In the limit that the relaxation and dephasing times are much longer than the gate time $t_{\rm g} = \pi/2g$, each dephasing or relaxation process contributes independently to the gate infidelity, i.e., $r_{\rm incoh}^{\rm iSWAP} = r_1 +r_2 +r_3 +r_4$, where $r_1(r_{2})$ is the error due to $\Gamma_{1,1}$($\Gamma_{1,2}$), and $r_{3} (r_{4}$) is the error due to $\Gamma_{\phi,1}$($\Gamma_{\phi,2}$). Let us first consider the coherent evolution plus the first qubit relaxation terms. The infidelity of the iSWAP gate due to the relaxation process is obtained by comparing the superoperator matrix $\mc{S}_{\rm iSWAP}$ with the ideal iSWAP superoperator matrix $u_{\rm iSWAP} = U_{\rm iSWAP}^{\dag}\otimes U_{\rm iSWAP}$. Substituting \cref{SiSWAP} and $u_{\rm iSWAP}$ into \cref{fidelity}, and expanding the resulting equation to leading order in $\Gamma_{1,1}$ using Mathematica, we obtain: 
\begin{align}
    r_{1}& = \frac{1}{20}[10 + 3 \Gamma_{1,1} t_{\rm g} + (2-\Gamma_{1,1}t_{\rm g}) \cos(
   2 g t_{\rm g} )\notag\\
   &+ 4( \Gamma_{1,1}t_{\rm g} -2) \sin(g t_{\rm g})].
\end{align}
Evaluating the unitary evolution (the oscillatory terms) at $t_{\rm g} = \pi/2g$, we find:
\begin{align}
    r_{1}= \frac{2}{5}\Gamma_{1,1}t_{\rm g}.
\end{align}
We intentionally remove the coherent error by evaluating the oscillatory terms at $\tau = \pi/2g$.
Repeating the above steps for the remaining relaxation and dephasing processes, we obtain the incoherent error  ($r_{\rm incoh.} = 1- 
%\bar{F}_{\rm P}$) 
\bar{F}_{\mc{E},\mc{U}_{\rm iSWAP}}$) for iSWAP to be
\begin{align}
\label{incoh_error}
    r_{\rm incoh} \approx \frac{2}{5}\sum_{k=1}^2(\Gamma_{1,k}+\Gamma_{\phi,k}^{\rm wh})t_{\rm g}.
\end{align}
Introducing $\Gamma_{\phi,k}^{\rm wh} = \Gamma_{2,k}-\frac{1}{2}\Gamma_{1,k}$, the incoherent error can be rewritten in terms of the relaxation rate $\Gamma_{1,k}$ and the total dephasing rate $\Gamma_{2,k}$ as
\begin{align}
    r_{\rm incoh} \approx \frac{2}{5}\sum_{k=1}^2\left(\frac{1}{2}\Gamma_{1,k}+\Gamma_{2,k}\right)t_{\rm g}.
\end{align}
Note that this error is only for the active part of the gate. If there is a waiting or padding time before and after the gate pulse, the error incurred due to the relaxation and dephasing processes will be the same except for replacing $t_{\rm g}$ with $t_{\rm w}$. In general, \cref{incoh_error} is valid for any gate that is implemented within the computational subspace. 

\section{Incoherent error for a CZ gate}\label{incoh_CZ}
Here we present the derivation of the approximate incoherent error for the CZ gate in the limits $\Gamma_1t_{\rm g}\ll1$ and $\Gamma_{\phi, \rm w}t_{\rm g} \ll 1$. A CZ gate between the two qubits is activated by coupling, for example, $|11\rangle$ with $|20\rangle$ with the Hamiltonian
\begin{align}
    H_{\rm CZ20} = g(|11\rangle\langle 20|+|20\rangle\langle 11|).
\end{align}
A full period evolution from $|11\rangle$ to $|20\rangle$ and back yields the conditional phase $e^{-i \pi}$ to $|11\rangle$ needed to define CZ. The dynamics of the system in the presence of relaxation and dephasing can be described by
\begin{align}\label{master_eqn_2}
 \dot \rho(t) &= -i[H_{\rm CZ20},\rho(t)] + \frac{1}{2}\sum_{k=1}^2\Gamma_{1,k}\mc{D}[\sigma_{k}]\rho(t) \notag\\
    &+\sum_{k=1}^2\Gamma_{\phi,k}^{\rm wh}\mc{D}[n_{k}]\rho(t)
\end{align}
where $a_{k} = (|0\rangle\langle 1| + \sqrt{2}|1\rangle \langle 2|)_{k}$ and $n_{k} = (|1\rangle\langle 1|+2|2\rangle\langle 2|)_k$. We again first find the superoperator matrix by exponentiation of $\mc{L}$ given by \cref{eq:L-mat}:
\begin{align}
    \mc{L} = & -i \mathbbm{1}\otimes H_{\rm CZ20} + i H_{\rm CZ20}^{T}\otimes\mathbbm{1}\notag\\
    &+\frac{1}{2}\sum_{k=1}^2\Gamma_{1,k}(2a_{k}\otimes a_{k}-\mathbbm{1}\otimes  n_{k}- n_{k}\otimes \mathbbm{1}) \\
    &+\sum_{k=1}^2\Gamma_{\phi,k}^{\rm wh}(2 n_{k}\otimes  n_{k} - \mathbbm{1}\otimes  n_{k}^2- n_{k}^2\otimes \mathbbm{1}),\notag
\end{align}
and projecting into the qubit subspace, i.e.:
\begin{align}
    \mc{S}_{\rm CZ} = Pe^{\mc{L}t_{\rm g}}P^{\dag},
\end{align}
where $P = p\otimes p$ with $p=p_{\rm proj}\otimes p_{\rm proj}$ and 
\begin{align}
%p_{\rm proj} = (1,0)\otimes (1,0,0)^{T} +(0,1)\otimes (0,1,0)^{T}.
p_{\rm proj} = (1,0)^T (1,0,0) +(0,1)^T (0,1,0).
\end{align}
In the limit that the relaxation and dephasing times are much longer than the gate time $t_{\rm g} = \pi/g$, each dephasing or relaxation process contributes independently to the gate infidelity, i.e., $r_{\rm incoh}^{\rm CZ20} = r_1 +r_2 +r_3 +r_4$. Let us first consider the coherent evolution plus the second qubit relaxation terms. The infidelity of the CZ gate due to the relaxation process is obtained by comparing the superoperator matrix $\mc{S}_{\rm CZ}$ with the ideal CZ superoperator matrix $u_{\rm CZ} = U_{\rm CZ}^{\dag}\otimes U_{\rm CZ}$.
\begin{align}\label{r1_CZ}
    r_{2} &= 1 -\bar{F_{2}}  \approx \frac{1}{10}\cos^2(gt_{\rm g}/2)[7-\cos(gt_{\rm g})]\\
    &+\frac{\Gamma_{1,2} }{40 g}\left\{  [3 - \cos(g t_{\rm g} )][2gt_{\rm g} - gt_{\rm g} \cos(g t_{\rm g} )- \sin(g t_{\rm g}) ]\right\}\notag
\end{align}
Evaluating the oscillatory terms in \cref{r1_CZ} at $t_{\rm g} = \pi/g$, we get the error due to the second qubit relaxation
\begin{align}
    r_{2} = \frac{3}{10}\Gamma_{1,2}t_{\rm g}.
\end{align}
Following the same reasoning, the contribution of relaxation of the first qubit is $r_{1} = \Gamma_{1,1}t_{\rm g}/2$. Note that since the first qubit oscillates between $|1\rangle$  and $|2\rangle$, the relaxation processes of the first qubit contribute more to the gate error than the second qubit that oscillates between $|0\rangle$ and $|1\rangle$ during the gate (assuming the same relaxation rate for the qubits). Computing the contribution of the dephasing processes, the total error of CZ due to relaxation and dephasing processes reads  
\begin{align}\label{cz_rate}
r_{\rm incoh}^{\rm CZ20} \approx  \frac{1}{2}\Gamma_{1,1}t_{\rm g} +\frac{3}{10}\Gamma_{1,2}t_{\rm g} +\frac{61}{80}\Gamma_{\phi,1}^{\rm wh}t_{\rm g}+\frac{29}{80}\Gamma_{\phi,2}^{\rm wh}t_{\rm g}.
\end{align}
It is worth noting that the contribution of the first qubit dephasing to the total error is stronger than that of the second qubit because the coherent oscillations for the first qubit occur between $|1\rangle$ and $|2\rangle$ and hence more dephasing comes from the state $|2\rangle$. The error rate \cref{cz_rate} only accounts for the flat part of the pulse. If there is padding before and after the pulse, the error is given by 
\begin{align}
r_{\rm incoh}^{\rm CZ20} &\approx  \frac{1}{2}\Gamma_{1,1}t_{\rm g} +\frac{3}{10}\Gamma_{1,2}t_{\rm g} +\frac{61}{80}\Gamma_{\phi,1}^{\rm wh}t_{\rm g}+\frac{29}{80}\Gamma_{\phi,2}^{\rm wh}t_{\rm g} \notag\\
&+\frac{2}{5}\sum_{l=1}^2\left(\Gamma_{1,k}+\Gamma_{\phi,k}^{\rm wh}\right)t_{\rm w}.
\end{align}
The incoherent error can be expressed in terms of the total dephasing rates $\Gamma_{2,k}$ and relaxation rates $\Gamma_{1,k}$ as
\begin{align}\label{error_CZ20}
r_{\rm incoh}^{\rm CZ20} &\approx  \frac{19}{160}(\Gamma_{1,1} +\Gamma_{1,2})t_{\rm g} +\frac{61}{80}\Gamma_{2,1}t_{\rm g}+\frac{29}{80}\Gamma_{2,2}t_{\rm g} \notag\\
&+\frac{2}{5}\sum_{l=1}^2\left(\frac{1}{2}\Gamma_{1,k}+\Gamma_{\phi,k}^{\rm wh}\right)t_{\rm w}.
\end{align}

A CZ gate can also be activated by the interaction Hamiltonian
\begin{align}
 H_{\rm CZ02} = g(|11\rangle\langle 02|+|02\rangle\langle 11|).
\end{align}
The corresponding incoherent error can be obtained by replacing $1\leftrightarrow 2$ in \cref{error_CZ20} 
\begin{align}\label{error_CZ02}
r_{\rm incoh}^{\rm CZ02} &\approx  \frac{19}{160}(\Gamma_{1,2} +\Gamma_{1,1})t_{\rm g} +\frac{29}{80}\Gamma_{2,1}t_{\rm g} +\frac{61}{80}\Gamma_{2,2}t_{\rm g} \notag\\
&+\frac{2}{5}\sum_{l=1}^2\left(\frac{1}{2}\Gamma_{1,k}+\Gamma_{\phi,k}^{\rm wh}\right)t_{\rm w}.
\end{align}

\section{Error due to $1/\mathrm{f}$ flux noise}
\label{one_on_f}

Let us consider dephasing due to $1/\mathrm{f}$ flux noise only, assuming both qubits are sensitive to flux noise. Assuming the qubit is susceptible to $1/\mathrm{f}$ flux noise only, the off-diagonal element of the qubit density matrix decays as $\sim \exp[-(\Gamma_{\phi}^{1/\mathrm{f}}t)^2]$, which can be modeled using the master equation

\begin{align}\label{master_one_on_f}
 \dot \rho(t) = -i[H,\rho(t)] + \sum_{k=1}^22 t \left(\Gamma_{\phi,k}^{1/ {\rm f}}\right)^2\mc{D}[n_{k}]\rho(t),
\end{align}
where  $\Gamma_{\phi,k}^{1/{\rm f}}$ are the dephasing rates of the qubits due to $1/\mathrm{f}$ flux noise. The vectorized master equation for the CZ gate activated with a coupling Hamiltonian $H = g(|11\rangle\langle 20|+|20\rangle\langle 11|)$  has the form
\begin{align}
    \dot{\vec{\rho}} (t) = e^{\int_{0}^{\tau} dt' \mc{L}(t')}\overrightarrow{\rho(0)},
\end{align}
where the time-dependent superoperator is given by 
\begin{align}
    \mc{L}(t) = & -i \mathbbm{1}\otimes H + i H^{T}\otimes\mathbbm{1}\notag\\
    &+\sum_{k=1}^22 t (\Gamma_{\phi,k}^{1/ {\rm f}})^2(t)(2 n_{k}\otimes  n_{k} - \mathbbm{1}\otimes  n_{k}^2- n_{k}^2\otimes \mathbbm{1}),
\end{align}
In the limit $\Gamma_{\phi,k}^{1/{\rm f}} \tau \ll 1$, and following the same procedure in Appendix \ref{incoh_CZ}, the error rate due to $1/\mathrm{f}$ flux noise reads~\cite{Didier2019}
\begin{align}\label{CZ20_one_on_f}
    r^{\rm CZ20}_{T_{\phi,1/\mathrm{f}}} = \frac{61}{80}\left(\Gamma_{\phi,1}^{1/{\rm f}} t_{\rm g}\right)^2 + \frac{29}{80} \left(\Gamma_{\phi,2}^{1/{\rm f}} t_{\rm g}\right)^2. 
\end{align}
The error rate for the CZ02 gate due to $1/\mathrm{f}$ flux noise can be obtained by interchanging the indices of the qubits in \cref{CZ20_one_on_f}. Similarly, for the iSWAP gate, the error due to $1/\mathrm{f}$ flux noise (assuming no coherent error) has the form 
\begin{align}
    r^{\rm iSWAP}_{T_{\phi,1/{\rm f}}} = \frac{13}{20} -\frac{1}{2}e^{-\Gamma^2 t_{\rm g}^2/2} -\frac{3}{20}e^{-\Gamma^2 t_{\rm g}^2}
\end{align}
where $\Gamma^2 = \left(\Gamma_{\phi,1}^{1/{\rm f}}\right)^2 + \left(\Gamma_{\phi,2}^{1/{\rm f}}\right)^2$. In the limit $\Gamma_{\phi,k}^{1/{\rm f}}t_{\rm g}\ll 1$, we have
\begin{align}
    r^{\rm iSWAP}_{T_{\phi,1/\mathrm{f}}} = \frac{2}{5}\left[\left(\Gamma_{\phi,1}^{1/{\rm f}}\right)^2 + \left(\Gamma_{\phi,2}^{1/{\rm f}}\right)^2\right] t_{\rm g}^2.
\end{align}

\clearpage
\bibliography{references.bib}
\end{document}